\documentclass[twocolumn]{article}
\usepackage{pgfplots}
\pgfplotsset{compat=newest}
\usepackage[utf8]{inputenc}
\usepackage{microtype}
\usepackage[numbers,square]{natbib}
\usepackage{authblk}

\usepackage{booktabs}
\usepackage{color}
\usepackage{url}
\usepackage{graphicx}
\usepackage{amsmath}

\usepackage{acronym,url}

\usepackage{makecell}
\usepackage{enumitem}
\usepackage{subcaption}

\usepackage{float}

\usepackage{soul}

\acrodef{IDE}[IDE]{Interactive Differential Evolution}
\acrodef{BAQ}[BAQ]{Basic Audio Quality}
\acrodef{JNDs}[JNDs]{Just Noticeable Differences}

\graphicspath{{./}{figures/}}
\usepackage{hyperref}
\usepackage{tabularx}

\title{An Adaptive Method for Target Curve Selection\\ \large \textbf{Preprint}}

\author[1]{Gabriele Ravizza}
\author[2]{Juli\'an Villegas}
\author[1]{Christer P.\ Volk}
\author[1]{Tore Stegenborg-Andersen}
\author[2]{Yan Pei}

\affil[1]{SenseLab, FORCE Technology, Hørsholm, Denmark}
\affil[2]{University of Aizu, Aizu Wakamatsu, Japan}

\usepackage[pscoord]{eso-pic}
\newcommand{\placetextbox}[3]{
\setbox0=\hbox{#3}
\AddToShipoutPictureFG*{
\put(\LenToUnit{#1\paperwidth},\LenToUnit{#2\paperheight}){\vtop{{\null}\makebox[0pt][c]{#3}}}%
}%
}%

\date{}

\begin{document}

\twocolumn[
\maketitle
\begin{abstract}
In this paper, we introduce an adaptation of the ``Interactive Differential Evolution'' (IDE) algorithm to the audio domain for the task of identifying the preferred over-the-ear headphone frequency response target among consumers. The method is based on data collection using an adaptive paired rating listening test paradigm (paired comparison with a scale). 
The IDE algorithm and its parameters are explained in detail. Additionally, data collected from three listening experiments with more than 20 consumers is presented, and the algorithm's performance in this untested domain is investigated on the basis of two convergence measures. The results indicate that this method can converge and may ease the task of 'extracting' frequency response preference from untrained consumers.
\end{abstract}
\vspace{12pt}
]

\placetextbox{0.5}{0.08}{\fbox{\parbox{\dimexpr\textwidth-2\fboxsep-2\fboxrule\relax}{\footnotesize \centering Accepted for presentation at the Audio Engineering Society (AES) International Conference on Headphone Technology, 2025 August 27–29, Espoo, Finland.}}}

\section{Introduction}
Headphones and earbuds have become increasingly popular for consuming most media content. 
Finding the best tuning strategies in an increasingly competitive market is paramount. 
Audio quality preference 
is idiosyncratic but has been shown to share some traits among certain groups of people \cite{Rentfrow2003, Anderson2020}. 
One challenge is how to best \textit{extract} a preferred frequency response curve from untrained consumers. 
Ideally, consumers should be consulted to obtain a representative result, but doing so limits the task that can be asked of them. 
Two main approaches have been used: 
Asking consumers to rate their preference of pre-selected candidate frequency response curves or physical benchmark products (see e.g., \cite{gabrielsson1990perceived} by Gabrielsson et al.) and 
asking consumers to adjust frequency bands to the most preferred curve. 
The first approach comes with the challenge of pre-selecting these candidate curves. 
To ensure an efficient listening test with high data quality, the number of curves to be evaluated must be limited. 
Moreover, each curve must be evaluated with a representative number of samples (musical excerpts), which poses a scalability problem.  
 Furthermore, there is a limited amount of literature on loudness JNDs 
(Just Noticeable Differences---the smallest detectable difference in sound level), especially for particular bands in broadband audio, such as music, hindering efficient selection of curves that are sufficiently distinct for discrimination by na\"ive, normal-hearing listeners.

Adjusting a frequency response curve is a difficult task for untrained listeners and can lead to sub-optimal curves. 
To accommodate for this challenge, Olive et al.\ \cite{olive2015factors} proposed a method where listeners are asked to adjust only two filters, one for bass and one for treble, in a so-called method of adjustments. While this eases the task, it also limits the scope of the test. 
Trained listeners might adapt to the task more easily, but they also pose a risk of introducing bias (from training), potentially leading to outcomes that don’t accurately reflect the intended target audience for the brand or product.

Our proposed method is inspired by both approaches to push the boundaries of listening test methodology. 
In this method, each listener's response is evaluated and used to generate stimuli for the following trial. 
This study builds upon knowledge acquired from a previous study from 2023, specifically investigating listening preference target curve selection for over-ear headphones measured on a B\&K Type 5128 \cite{SenseLabAizu2023}, where a set of preferred curves was found. 
This set has informed the starting point to test our newly proposed method. 
\section{Methods}
\subsection{Study Design}
\begin{figure}[ht!]
\begin{center}
\includegraphics[width=0.41\textwidth, viewport = 0 0 499 424]{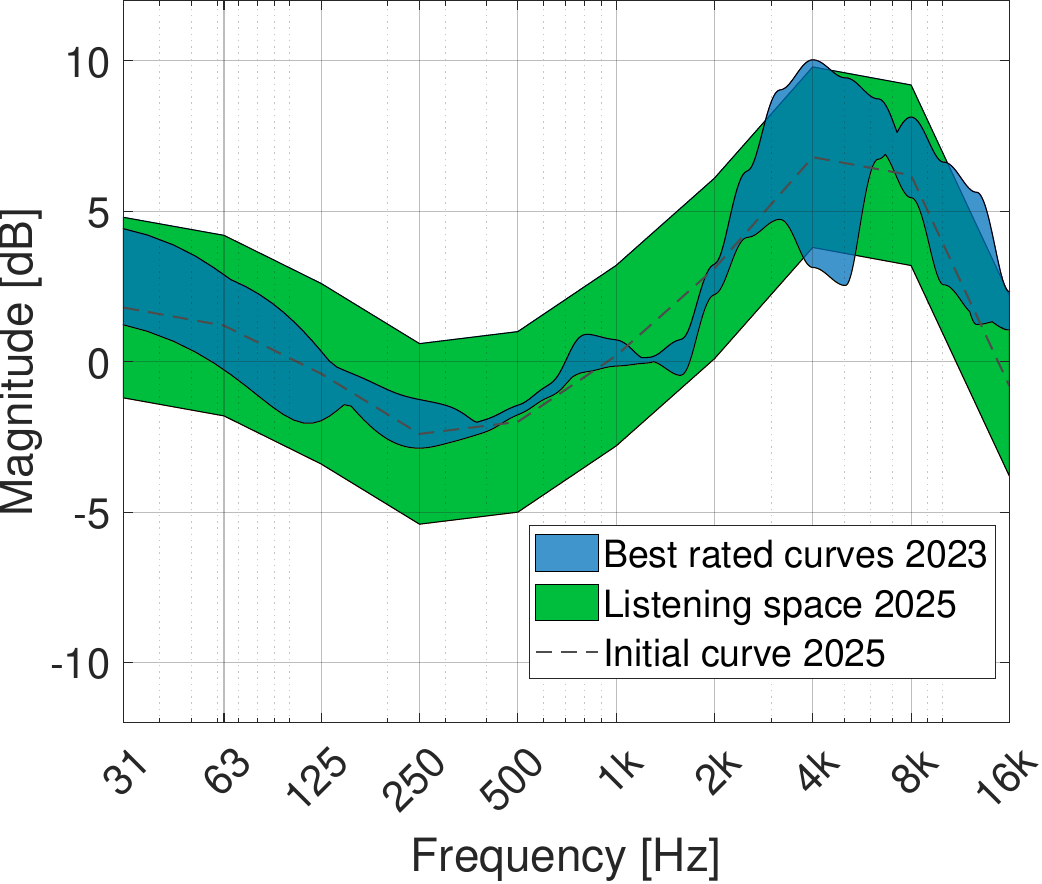}
\caption{Listening space selected in the current research (green) compared to the most preferred curves in previous research (blue) \cite{SenseLabAizu2023}. The initial curve is the median of the green area. 
}
\label{fig:listening_space}
\end{center}
\end{figure}

We use an interactive evolutionary computation algorithm \cite{Pei2023}, Interactive Differential Evolution (IDE), to optimize a population of target frequency response curves through comparison. 
The method was originally developed to simulate random gene mutation and biological evolution \cite{storn1997differential}.
Each ``individual candidate'' within this population represents a potential target curve, effectively an array of gains (genes) for different frequency bands. 
By incorporating subjective human auditory ratings as the fitness measure, the method effectively guides the optimization process toward perceptually optimized frequency response curves.
These gains are randomly initialized within specified bounds, inspired by the previous study mentioned earlier \cite{SenseLabAizu2023}. 
These bounds were selected to span the range of the best-rated curves from \cite{SenseLabAizu2023}, and modified (mostly expanded) to be within a uniform range of $\pm3$\,dB as shown in \autoref{fig:listening_space}.
The population is then progressively evolved by applying the operations of mutation (at a rate $F$, i.e., the scaling factor) and crossover (at a rate $C$). See details in \autoref{sec:IDE_theory}.

The specific bounds found in \cite{SenseLabAizu2023} were obtained by having consumers evaluate the frequency responses of $32$ curves inspired by existing headphones, resulting in a frequency response range hypothesized to include the preferred consumer target.
However, this method had limitations, particularly regarding generalizing the spectral similarities between the top-rated curves. 
In \cite{olive2015factors}, headphone responses were shaped using filters, enabling adjustments to the bass and treble frequencies.  
Our method extends upon that approach by shaping the target response curve candidates across ten one-octave frequency bands, ranging from $31$ to $16$k\,Hz.

The experiment was divided into a comparison stage and an evaluation stage, in that order. 
In comparison trials, two curves were pitted against each other: a reference and a modified reference curve. 
Participants rated the pitted curves on a continuous bipolar rating scale according to their preference, and the most preferred curve (binary selection criteria) was selected by the algorithm for further evolution. 
This process was repeated over multiple generations until a prescribed number of generations was reached. 
For the experimental design, the algorithm facilitated a sequence of trials with a randomized presentation order, and the reference and mutation were randomly switched in each trial to reduce order effects.  

In the evaluation stage, participants compared the final generation population candidate curves against each other for each music sample. 
The use of randomization allowed for variability in trial conditions and helped avoid systematic biases in participant responses. 
Additionally, random presentation order of the music samples ensured that participants were exposed to a range of stimuli in a non-repetitive manner.

A process for dynamic population updating was implemented as well. 
The population was updated after each set of comparison trials, with individual candidates replaced by those preferred by the participant in each trial. 
A set comprised the population pool for a given generation.
Each time the population was updated, the new population was logged, and the current state of the experiment was saved to ensure that the study could be resumed when interrupted. 
This mechanism also allowed for tracking the population's evolution over time and evaluating the evolutionary computation algorithm's effectiveness in optimizing the curves.

In the listening experiment, the curve population consisted of five individuals (optimized for perceptual discrimination) and evolved through eight generations of interactive optimization. 
This design adheres to Miller's $7\pm2$ cognitive limit, ensuring listeners could effectively compare and evaluate auditory stimuli without overwhelming sensory memory \cite{miller1956magical} in the multiple-stimulus evaluation experiment.
\subsection{Interactive Differential Evolution}\label{sec:IDE_theory}
The interactive evolutionary computation algorithm used in this study employs two key parameters: 
the scaling factor $F$  and the crossover rate $C$ \cite{storn1997differential} (or crossover probability). 
These parameters control how individuals in the population evolve across generations. 
The scaling factor $F$ defines the magnitude of the mutation applied to the population, while the crossover rate $C$ determines the probability with which a gene (a frequency band) from the mutant vector is exchanged with the reference individual during the crossover operation.

To incorporate perceptual preferences into the optimization loop  \cite{Wang2024}, human listeners evaluated audio samples rendered from candidate frequency responses (individuals) through controlled listening tests. 
Each trial involved pairwise comparisons between the target and mutant vectors, where participants selected the perceptually preferred response \cite{Takagi}. 
These binary choices served as the fitness function, steering the evolutionary process toward subjectively optimal solutions. 
This interactive framework allows real-time adaptation of $F$ and $C$ based on listener feedback, for instance, reducing $F$ to fine-tune mutations when nearing perceptually stable regions.
However, these parameters were kept constant in our implementation.
\subsubsection*{Mutation Operation}
%
%
Differential evolution's mutation creates an individual ($\mathbf{m}_i$) by adding a scaled difference ($F$) between randomly selected population individuals ($\mathbf{b}_k- \mathbf{c}_l$) to a base individual ($\mathbf{a}_j$), promoting exploration of new regions in the search space.
Specifically, a mutant individual is generated as
\[
\mathbf{m}_i = \mathbf{a}_j + F \left( \mathbf{b}_k- \mathbf{c}_l \right),
\]
where
\begin{itemize}
  \item \( \mathbf{m}_i \) is the mutated individual at position \( i \),
  \item \( \mathbf{a}_j, \mathbf{b}_k, \mathbf{c}_l \) are selected individuals from the population at randomly selected positions \( j, k, l \),
  \item \( F \) is the scaling factor, a constant that controls the intensity of the mutation.
\end{itemize}

The scaling factor $F$ is set between $0$ and $2$ with higher values introducing larger mutations.
In the listening experiment, the mutation factor $F$ was set to $0.2$.
\subsubsection*{Crossover Operation}
The crossover operation is performed between the reference individual and the mutant one to create a new trial individual. 
The crossover blends the mutant vector with the reference vector to retain some of the original information, working together with mutation to balance global exploration and local exploitation.
The crossover is done on a per-gene basis, with a probability $C$ for each gene being swapped between the reference and mutant individual as
\[
\mathbf{t}_i = 
\begin{cases} 
\mathbf{m}_i & \text{if }\ \text{random}(0,1) < C, \\
\mathbf{r}_i & \text{otherwise}, 
\end{cases}
\]
where
 \( \mathbf{t}_i \) is the trial individual at gene \( i \),
 \( \mathbf{r}_i \) is the reference individual at gene \( i \),
\( \mathbf{m}_i \) is the mutated individual at gene \( i \), and 
  \( C \) is the crossover rate, the probability of selecting the gene from the mutant.

If a randomly generated number is \(<C\), the trial individual inherits the gene from the mutant individual; otherwise, it inherits the gene from the reference individual.
In the listening experiment, the crossover rate \( C = 0.7\).

\subsubsection*{Bounds Handling}
After the crossover, the trial individual must be ensured to respect the predefined bounds for each band. 
This is done by clipping each gene to stay within its lower and upper limits:
\[
\mathbf{t}_c = \max\left(\min\left(\mathbf{t}_i, \mathbf{u}_i\right), \mathbf{l}_i\right)
\]
where 
\( \mathbf{t}_c \) is the clipped trial individual at gene \( i \), and
  \( \mathbf{l}_i \) and \( \mathbf{u}_i \) are the lower and upper bounds for the gene 
  \( i \), respectively.

\begin{figure}[htbp]
\begin{center}
\includegraphics[width=0.475\textwidth, viewport = 0 28 603 559]{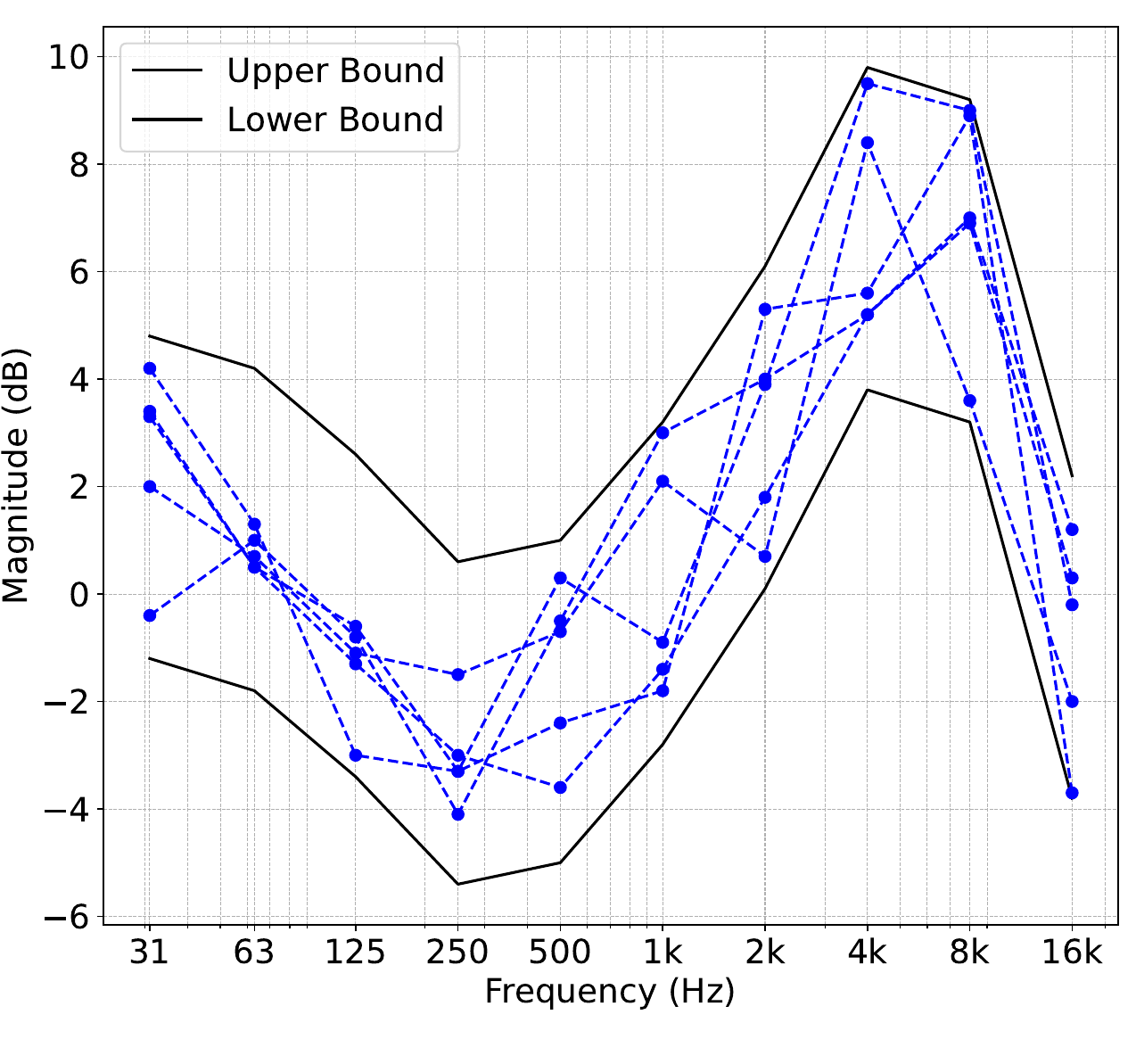}
\caption{Example of five candidate curves comprising a ``population'' within the bounds of the restricted space.}
\label{fig:population_space}
\end{center}
\end{figure}

An example of the resulting mutation curves is shown in \autoref{fig:population_space}.
\subsubsection*{Human Subjective Evaluation}
After performing mutation and crossover operations, the IDE algorithm compares the auditory characteristics of the target individual ($\mathbf{t}_i$) and trial vector ($\mathbf{r}_i$) through subjective human evaluation. 
This enables the method to progressively converge toward the perceptually optimal zone for human hearing. 
Through iterative generations, the optimization process refines the frequency response curves until they align with human auditory preferences.

\section{Listening Test}
\begin{figure}[ht!]
\begin{center}
\includegraphics[width=0.4\textwidth]{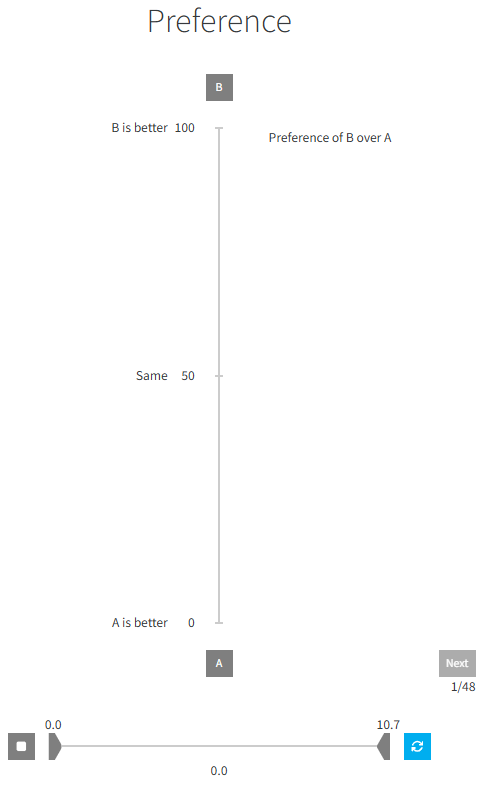}
\caption{Screenshot of the paired comparison (Experiment 1) interface in SenseLabOnline \cite{SenseLabOnline}.
}
\label{fig:senselabonline_exp1}
\end{center}
\end{figure}

\begin{figure}[ht!]
\begin{center}
\includegraphics[width=0.478\textwidth]{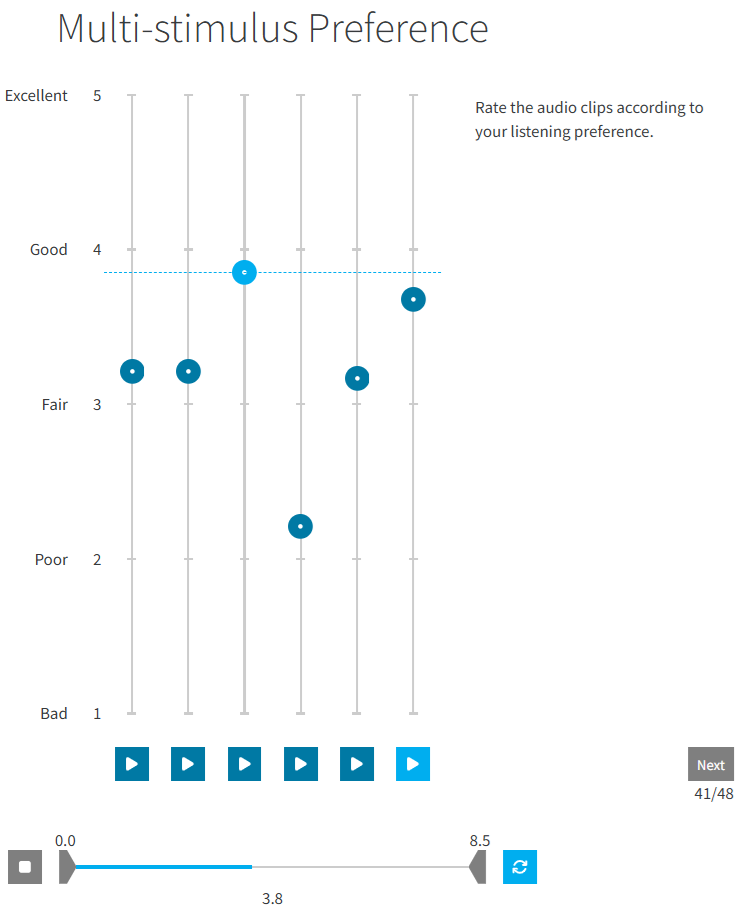}
\caption{Screenshot of the multiple stimulus evaluation (Experiment 2) interface in SenseLabOnline \cite{SenseLabOnline}. With the five final candidate curves and the flat response anchor.
}
\label{fig:senselabonline_exp2}
\end{center}
\end{figure}

The listening test consisted of two parts; a pairwise comparison with a rating scale---the main experiment, where the adaptive method described above was active (see \autoref{fig:senselabonline_exp1}), followed by a multiple stimulus test where the 5 individuals (curves) in the final population are rated against each other (see \autoref{fig:senselabonline_exp2}) together with a sixth flat curve included as an anchor for stabilizing the scale used.

The rating scale in the paired rating experiment was a continuous bipolar scale with three verbal anchors: ``A is better'', ``Same'', ``B is better'', respectively. Ratings on this scale were interpreted as simple binary preference by the algorithm, but the scale was chosen for diagnostic purposes.

In the multiple-stimulus evaluation experiment, the continuous $[1,5]$ rating scale from the ``Overall Quality'' attribute described in the ITU-T P.835 recommendation \cite{ITU-T:P835} was utilized, having five verbal anchors evenly distributed: $1$ (``Bad''), $2$ (``Poor''), $3$ (``Fair''), $4$ (``Good''), and $5$ (``Excellent'').
%
\begin{table}[h!tbp]
\begin{tabular}{@{}ll@{}}
\toprule
\textbf{Track title} & \textbf{Artist}          \\ 
\midrule
Fast Car             & Tracy Chapman            \\
Change The World     & Eric Clapton             \\
7 Years              & Lucas Graham             \\
Dreams               & Fleetwood Mac            \\
Get Lucky            & Daft Punk ft.\ Pharrell   \\
Hotel California     & The Eagles               \\
The Meaning of Blues & Claire Martin            \\
Know Who You Are     & Pharrell ft.\ Alicia Keys\\
\bottomrule
\end{tabular}
\caption{Listening test tracks included in the experiment. Excerpts were 15-30 seconds long.}
\label{tab:test_tracks}
\end{table}

Audio was presented over Beyerdynamic DT770 Pro $250$\,$\Omega$ closed-back headphones. 
Playback, scales, randomization, data collection, and instruction presentation were configured in and handled by SenseLabOnline 5.4. \cite{SenseLabOnline} developed by FORCE Technology, with which some of this paper's authors are affiliated. 

In SenseLabOnline, a ``scripted test'' can be configured to use Python scripting to define the logic of a listening test. 
Via this tool, the algorithm was implemented, also utilizing a self-developed custom Python filter toolbox uploaded to the platform, which expressed the $10$ frequency bands as FIR filters, and in a post-filtering step, performed automatic loudness alignment according to ITU-R BS.1770-3\cite{ITU-R:BS1770-3}, set to $-18$ LUFS. 
The script allowed fast stimuli-generation between trials based on the current trial's rating(s), and additionally, the platform supports real-time filtering within each trial for compensation of the playback headphone (also during cross-fade between stimuli). The script also saved the parameters of each generated curve in each trial.
\subsection{Playback headphone compensation filter design}
The presentation headphones 
were measured using a B\&K HATS 5128C and SoundCheck v.22 (Listen, Inc.) with a logarithmic stepped sweep tone.
Four separate DT770 units were tested, with each unit repositioned three times during measurement.
No outliers, such as leaking positions, were observed.
The resulting average response included contributions from both the HATS 5128 and the DT770 headphones.
This average was then used to derive an inverse filter to compensate for the combined effects of the HATS and the headphones.
Sharp notches and peaks in the inverse filter were smoothed, and its frequency range was constrained to $16$Hz–$16$kHz.
The inverse filter was created and applied to the audio clips using the AKtools toolbox \cite{AK_tools}

\subsection{Participants \& test duration}
Twenty-four Japanese students participated in the comparison- \& evaluation experiment, while 22 participated in a follow-up benchmark experiment (introduced in \autoref{sec:convergence}).  
All participants were adults ($18$ years or older) and reported having normal hearing. No further requirements were set for the participants.

This research was conducted under the institutional ethics framework and following the University of Aizu’s procedures.
The combined test duration of the main comparison \& evaluation experiments ranged from $6.5$ minutes to $50$ minutes with an average of $31.5$ minutes (median of $\sim 31$ minutes).

\section{Results}
\label{sec:convergence}
To test the performance of the Differential Evolution algorithm, two measures are appropriate for direct evaluation:
\begin{enumerate}
    \item The standard deviation of the population pool of each generation
    \item Consumer rating of the preference of the best-ranked curve vs.\ the initial curve (dashed line in \autoref{fig:listening_space})
\end{enumerate}
The first measure is based on the hypothesis that the first-generation population, resulting from random mutations of the initial curve, will exhibit a larger variation among the pool of curves than subsequent generations. 
Assuming that consumers could perceive the difference between sets of competing curves and that they have a stable internal ``ideal''. Both might be true after an initial learning phase, but in this experiment, they had a limited number of generations for the curves to evolve sufficiently. 
One characteristic of our implementation of the IDE algorithm that is worth noting is that the size of its mutations across generations only decreases if the population curves converge, since we kept $F$ fixed. With an inconsistent input, mutations would be as big in the first generations as in the last; i.e., convergence is not inherent.

\begin{figure}[htbp]
\begin{center}
\includegraphics[width=0.45\textwidth, viewport = 5 5 319 294]{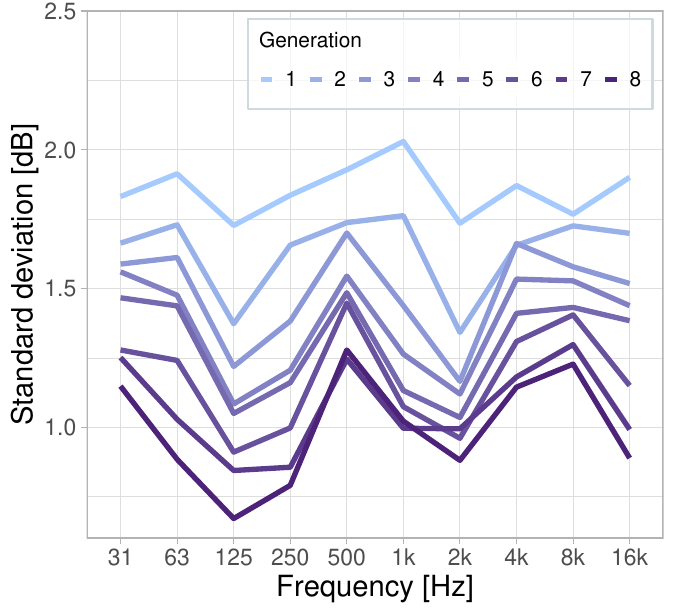}
\caption{Standard deviation of the pool of curves per generation and one-octave frequency band. 
Standard deviations were calculated on the basis of the eight curves in the population pool per assessor and then averaged.}
\label{fig:convergence_sd}
\end{center}
\end{figure}
In \autoref{fig:convergence_sd}, the averaged standard deviation across assessors of the pool of curves per generation and one-octave frequency bands is shown.
The purpose of investigating this per frequency band was to understand whether convergence occurred in all bands (if at all) or whether some bands were more difficult. 
The figure shows a clear decrease in standard deviation per generation. 
While some frequency bands decrease less (e.g., $500$\,Hz and $8$\,kHz), the trend is clear in all bands. 
Averaged across bands, the standard deviation decreases monotonically from $1.72$ to $0.98$. 
This decrease 
becomes smaller with each generation. 
It could be hypothesized that when a generation has not led to a reduction in standard deviation, the algorithm has converged to a level where consumers are no longer sensitive enough (due to discrimination ability or internal target stability) or disagree.

\begin{figure}[htbp]
\begin{center}
\includegraphics[width=0.45\textwidth, viewport =  5 23 283 287]{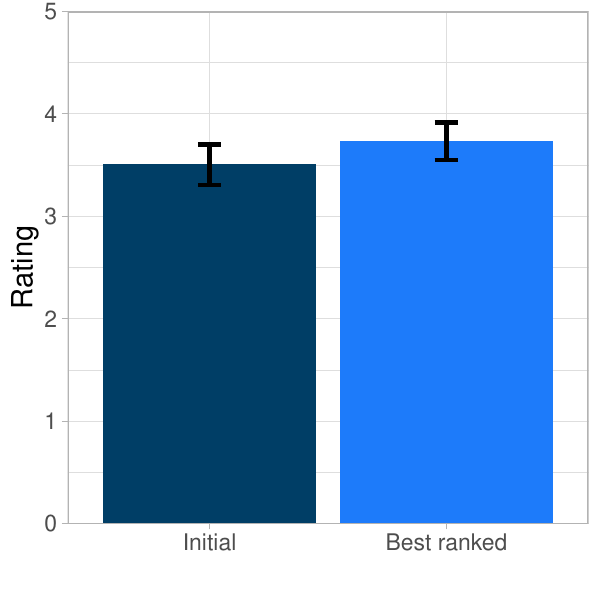}
\caption{Rating of the initial curve and the evolved averaged best-ranked curve. Error bars are $95\,\%$ confidence intervals.}
\label{fig:convergence_pref}
\end{center}
\end{figure}

The second measure directly investigates whether the preference for the best-ranked curve is higher than that of the initial curve. This tests whether the IDE converged on a reasonable target curve or something arbitrary. 
The initial curve (median from the first study \cite{SenseLabAizu2023}) was selected as the benchmark curve, as the average of an increasing number of random curves across all first-generation pools and assessors is expected to approach this initial curve. 
Preference assessments were analyzed with generalized linear mixed models \cite{glmmTMB}. The final model included the effect of System (`Initial' or `Best ranked'), a random intercept per assessor, and a random slope for System per assessor. The effect of the listening test track was not significant $[\chi^2(7)=9.974, p=.190]$. The difference between levels of the System was analyzed using estimated marginal means. 
This analysis showed that the odds of `Best ranked' to be preferred were $1.29$ times those for Initial [$z = 2.72, p=.018$], as illustrated in \autoref{fig:convergence_pref}.

The small difference between the curves shown in \autoref{fig:convergence_pref} is either a sign of limited convergence or that the initial curve was perceptually close to the preferred frequency response target (as would be expected).

%
\section{Final remarks}
We introduced the ``Interactive Differential Evolution'' (IDE), an algorithm inspired by gene mutations in biology. 
The algorithm was tested in the difficult task of using untrained consumers to find an ideal target curve for headphone reproduction. 
A main experiment with $24$ consumers was analyzed in an effort to evaluate the convergence performance of the algorithm. 
The results showed that for each ``generation'' of the experiment, the standard deviation of candidate curves' frequency bands decreases. 
Averaged across all frequency bands, the decrease over eight generations was monotonic, starting at $1.85~dB$ and ending at $0.99~dB$. 
This is a remarkable result considering that listeners were indirectly making changes in $10$ frequency bands, compared to e.g., the two bands used in the method of adjustments proposed by Olive et al. \cite{olive2015factors}. Although the two methods are not directly comparable, the end goal is.

The third experiment with $22$ consumers showed that the best-ranked of the evolved curves from the IDE was significantly preferred over the starting point, the initial curve (from the 2023 paper \cite{SenseLabAizu2023}), although only by a small margin (mean ratings of $3.74$ and $3.51$, respectively). 
Note that the frequency response of the best-ranked curve was not presented, as that was not the main focus of this study.

Since the experiment only included eight generations of evolution and was not very long (mean of $31.5$ minutes), the pool of candidate curves could potentially continue to converge, although the lack of knowledge of \acrodef{JNDs} for realistic stimuli limits assessment of how close our competing curves were to the sensitivity limits of untrained listeners. 
Knowing the \acrodef{JNDs} could also be particularly useful for choosing design target- and production tolerances of equal preference for normal hearing consumers.

While we did test the IDE algorithm with some different values of the mutation factor and crossover, there is probably still room for improvement. 
Changing these values seemed to have a big influence. More experiments are needed to understand the potential of IDE in perceptual audio evaluation, but this study indicates that it might be worth doing.

\section{Acknowledgments}
We thank Mr.\ W.\ Wen for administering the experiments, all the participants, and the University of Aizu for the competitive funding granted
for this research.

\bibliographystyle{plain}
\begin{small}
\setlength{\bibsep}{0pt plus 0.3ex}
\bibliography{refs}
\end{small}

\end{document}